\documentstyle[preprint, aps]{revtex}

\draft
\tighten

\begin{document}
\title{Dibaryon systems in the quark mass density- and temperature-dependent model}
\author{Yun Zhang$^{1,3}$ and Ru-Keng Su$^2$}
\address{$^1$Department of physics, Fudan University, Shanghai 200433 , P. R. China\\
$^2$China Center of advanced Science and Technology (World Laboratory)\\
P. O. Box 8730, Beijing 100080, P. R. China \\
$^3$Surface Physics Laboratory (National Key Laboratory), Fudan University,\\
Shanghai 200433 , P. R. China }
\maketitle

\begin{abstract}
Using the quark mass density- and temperature-dependent model, we have
studied the properties of the dibaryon systems. The binding energy, radius
and mean lifetime of $(\Omega \Omega )_{0^{+}}$and $(\Omega \Xi ^{-})$ are
given. We find the dibaryons $(\Omega \Omega )_{0^{+}}$, $(\Omega \Xi ^{-})$%
, $(\Omega \Xi ^0)$ are metastable at zero temperature, but the strong decay
channel for $(\Omega \Omega )_{0^{+}}$ opens when temperature arrives at $%
129.3$ $\mbox{MeV}$. It is shown that our results are in good agreement with
those given by the chiral $S(3)$ quark model.
\end{abstract}


\section{Introduction}

Since Jaffe predicted the existence of $H$ particle \cite{Jaffe}, the study
of dibaryon systems has attracted more and more attentions \cite
{su2,su3,su4,d,z1,z3,z2,su9,su10,su11,su12,su13,su14}. A lot of theoretical
models, for example, the constituent quark model in which the effective
degrees of freedom are constituent quarks and gluons \cite{su12}; the
quark-mesons exchange model in which the effective degrees of freedom are
constituent quarks and Goldstone bosons \cite{su9,su13}; the chiral SU(3)
quark model \cite{z1,z3,z2,su14}, ... had been employed to predict the bound
states of dibaryon systems. Many particles, namely, $H$ dihyperon \cite
{Jaffe,su2,su3}, $d^{\prime }$ particle \cite{d}, $(\Omega \Omega )_{0^{+}%
\text{ }}$and $\Omega ^{-}\Xi ^0$ \cite{z1,z3,z2,su14}, $\Delta \Delta $ 
\cite{su9,su11} have been suggested to be candidates for being observed in
the experiment. Unfortunately, the convincing experimental evidence for the
existence of these particles has not yet been found. Then it is essential to
confirm which dibaryon would be the best one to be detected in the
experiment. Therefore, it is necessary to address the above dibaryons from
different point of views, different models and different treatments, because
it can open the field of vision and choose the best detectable candidate
after comparing the results given by different models. This is the first
motivation to employ our quark mass density- and temperature-dependent
(QMDTD) model \cite{ourpaper,9,10,mpl} to study the dibaryon system.

The second motivation is to extend the study of dihyperons to finite
temperature. The strangeness enhancement was observed in relativistic heavy
ion collision (RHIC) recently \cite{su18,su19}. The dihyperon can be found
from the extreme condition of RHIC easily than that from the usual one.
According to the thermal model of RHIC, a fireball with high temperature and
high density is formed. Employing a suitable model to study the stability
and the thermodynamic properties of dibaryon at finite temperature can help
us to learn the detectable possibility of dihyperon.

In previous papers, by means of the QMDTD\ model, we studied the
thermodynamic behavior and stability of $A=5$ and $A=10$ strangelets at
finite temperature where $A$ is the baryon number \cite{ourpaper,mpl}. We
found that the stable strangelets only exist in the high strangeness module
and high negative charge region. For $A=2$ dibaryon system, $(\Omega \Omega
)_{0^{+}\text{ }}$has the maximum strangeness module and maximum negative
charge. It is of interest to study $A=2$ dibaryon system by using QMDTD
model.

The organization of this paper is as follows. In next section, we give the
main formulas of the QMDTD model. The results including stability, binding
energy, radius, lifetime for dibaryon systems at zero temperature are
presented in Sec. III. In Sec. IV, we will discuss the stability of dibaryon
via possible strong and weak decays at finite temperature. The last section
contains a summary.

\section{QMDTD model}

QMDTD model is a non-permanent quark confinement model \cite{9} because it
is based on the Friedberg-Lee model \cite{su20}. It has been used to study
the thermodynamic properties of strangelets \cite{ourpaper,10,mpl} and the
photo strange star \cite{su21}. The detail of this model can be found in 
\cite{ourpaper,9,10,mpl}. Here we only write down the main steps which is
necessary for calculating the thermodynamic quantities.

According to the QMDTD\ model, the masses of $u,$ $d$ quarks and strange
quarks (and the corresponding anti-quarks) are given by \cite
{ourpaper,9,10,mpl} 
\begin{eqnarray}
m_q &=&{\frac{B(T)}{3n_B}},\hspace{0.8cm}(q=u,d,\bar{u},\bar{d}),
\label{su1} \\
m_{s,\bar{s}} &=&m_{s0}+{\frac{B(T)}{3n_B}},  \label{su2}
\end{eqnarray}
where $n_B$ is the baryon number density, $m_{s0}$ is the current mass of
the strange quark and $B(T)$ is the vacuum energy density which satisfies 
\begin{eqnarray}
B(T) &=&B_0\left[ 1-a\left( \frac T{T_c}\right) +b\left( \frac T{T_c}\right)
^2\right] ,\text{ }0\leq T\leq T_c,  \label{15} \\
B(T) &=&0,\text{ }T>T_c,  \label{15-1}
\end{eqnarray}
where $B_0$ is the vacuum energy density inside the bag (bag constant) at
zero temperature , $T_{c\mbox{
}}=170$ $\mbox{MeV}$ is the critical temperature of quark deconfinement
phase transition, and $a$, $b$ are two adjust parameters. The values of $a,$ 
$b$ are fixed as 
\begin{equation}
a=0.65,\text{ }b=-0.35.  \label{18}
\end{equation}

Using the same argument as that of Ref. \cite{z1}, in RHIC\ experiments, the
fireball expands until freeze-out. The production of a large amount of
particles and violent collisions between them in the fireball finally lead
to the thermal and chemical equilibrium, soon after it reaches high
temperature and high density. The thermodynamic potential in the fireball
reads \cite{z1}

\begin{equation}
\Omega =-\sum_iT%
\displaystyle \int %
_0^\infty dk{\frac{dN_i(k)}{dk}}\ln \left( 1+e^{-\beta (\varepsilon
_i(k)-\mu _i)}\right) ,  \label{omiga}
\end{equation}
where $i$ stands for $u,$ $d,$ $s$ (or $\bar{u},$ $\bar{d},$ $\bar{s}$ )
quarks, $\mu _i$ is the corresponding chemical potential (for antiparticle $%
\mu _{\bar{i}}=-\mu _i$). $\varepsilon _i(k)=\sqrt{m_i^2+k^2}$ is the single
particle energy and $m_{i\text{ }}$is mass for quarks and antiquarks. ${%
{\displaystyle {dN_i(k) \over dk}}%
}$ is the density of states\ for various flavor quarks. For a dibaryon, its
geometry can be taken as a sphere. The density of states of a spherical
cavity has been calculated in our previous paper, it reads \cite{ourjpg}:

\begin{equation}
N_i(k)=A_i(kR)^3+B_i(kR)^2+C_i(kR),  \label{su6}
\end{equation}

\begin{equation}
A_i=\frac{2g_i}{9\pi },  \label{su7}
\end{equation}

\begin{equation}
B_i=\frac{g_i}{2\pi }\left\{ \left[ 1+\left( \frac{m_i}k\right) ^2\right]
\tan ^{-1}\left( \frac k{m_i}\right) -\left( \frac{m_i}k\right) -\frac \pi 2%
\right\} ,  \label{su8}
\end{equation}

\begin{equation}
C_i=\frac{g_i}{2\pi }\left\{ \frac 13+\left( \frac k{m_i}+\frac{m_i}k\right)
\tan ^{-1}\frac k{m_i}-\frac{\pi k}{2m_i}\right\} +\left( \frac{m_i}k\right)
^{1.45}{\frac{g_i}{3.42\left( {\frac{\displaystyle m_i}{\displaystyle k}}%
-6.5\right) ^2+100},}  \label{su9}
\end{equation}
where $g_i$ is the total degeneracy.

Since the quark mass $m_i$ depends on the density, the thermodynamic
potential $\Omega $ is not only a function of temperature, volume and
chemical potential, but also of density. The expression of total energy
density reads \cite{ourpaper,9,10,mpl}

\begin{equation}
\varepsilon ={\frac \Omega V}+\sum_i\mu _in_i-{\frac TV}\left. {\frac{%
\partial \Omega }{\partial T}}\right| _{\mu _i,n_B},  \label{18-2}
\end{equation}
where $n_B=A/V$ is the baryon number density and $A$ is the baryon number of
the strangelet, $V=%
{\displaystyle {4 \over 3}}%
\pi R^3$ is the volume of the strangelet. The number density of each
particle can be obtained by means of 
\begin{equation}
n_i=-{\frac 1V}\left. {\frac{\partial \Omega }{\partial \mu _i}}\right|
_{T,n_B}.  \label{19}
\end{equation}
At finite temperature, we must include the contributions of the
anti-particles, therefore, the baryon number for $i$ quark is given by 
\begin{equation}
\Delta N_i=(n_i-n_{\bar{i}})\times V=%
\displaystyle \int %
_0^\infty dk{\frac{dN_i(k)}{dk}}\left( \frac 1{\exp [\beta (\varepsilon
_i-\mu _i)]+1}-\frac 1{\exp [\beta (\varepsilon _i+\mu _i)]+1}\right) .
\label{20}
\end{equation}

The strangeness number $|S|$ of the strangelet reads 
\begin{equation}
|S|=\Delta N_s,  \label{21}
\end{equation}
the baryon number $A$ of the strangelet satisfies 
\begin{equation}
A={\frac 13}(\Delta N_u+\Delta N_d+\Delta N_s)\hspace{0in}.  \label{22}
\end{equation}
The electric charge $Z$ of the strangelet is 
\begin{equation}
Z=\frac 23\Delta N_u-%
{\displaystyle {1 \over 3}}%
\Delta N_d\hspace{0in}-%
{\displaystyle {1 \over 3}}%
\Delta N_s.  \label{23}
\end{equation}
At finite temperature, the stability condition of strangelets for the bag
radius $R$ reads 
\begin{equation}
\frac{\delta F}{\delta R}=0.  \label{13}
\end{equation}
where the free energy $F$ of strangelet is 
\begin{equation}
F=E-T\tilde{S},  \label{13-1}
\end{equation}
$E=\varepsilon V$ is the total energy, and 
\begin{equation}
\tilde{S}=\sum_i\tilde{S}_i=-\sum_i\left. {\frac{\partial \Omega }{\partial T%
}}\right| _{\mu _i,n_B}  \label{13-2}
\end{equation}
is the entropy. For the dibaryon system, baryon number $A$, the strangeness
number $|S|$, and electric charge $Z$ are given. Solving above equations
self-consistently, we obtain the thermodynamic quantities of dibaryon
systems.

\section{Result at zero temperature}

\subsection{Stability}

Firstly, we use the same arguments as that of Ref. \cite{8,ourpaper} to
study the stability of dibaryons at zero temperature. For a dibaryon $%
Q(A,|S|,Z)$, its baryon number $A$, strangeness number $|S|$ and electric
charge $Z$ are conserved in the strong process. A general expression of a
two-body strong baryon decay for a dibaryon bound state can be written as 
\begin{equation}
Q(2,|S|,Z)\rightarrow x(1,|S_x|,Z_x)+y(1,|S|-|S_x|,Z-Z_x),  \label{28}
\end{equation}
where $x$ and $y$ stand for baryons. For example, in the case of $(\Omega
\Omega )_{0^{+}\text{ }}$, $x$ and $y$ are free $\Omega $, respectively. At
zero temperature, the process (\ref{28}) is allowed if the energy balance of
the corresponding reaction satisfies 
\begin{equation}
E(2,|S|,Z)>m_x+m_y,  \label{29}
\end{equation}
where $E$ stands for the energy of the dibaryon bound state.

In the weak process, the baryon number $A$ and the electric charge $Z$ are
conserved, but the strangeness number $|S|$ is not conserved. For the weak
decay, $\Delta |S|=\pm 1$. Therefore, a general expression of a two-body
weak decay for dibaryon can be written as 
\begin{equation}
Q(2,|S|,Z)\rightarrow x(1,|S_x|,Z_x)+y(1,|S|-|S_x|\pm 1,Z-Z_x),.  \label{30}
\end{equation}
This process is allowed at zero temperature if the energy balance of the
corresponding reaction satisfies 
\begin{equation}
E(2,|S|,Z)>m_x+m_y,  \label{31}
\end{equation}
For example, the two-body weak decay to pure hadrons channel for $(\Omega
\Omega )_{0^{+}\text{ }}$is 
\begin{equation}
(\Omega \Omega )_{0^{+}\text{ }}\rightarrow \Omega +\Xi ^{-}.  \label{a1}
\end{equation}
In addition, $(\Omega \Omega )_{0^{+}\text{ }}$ can also decay to another
dibaryon and a hadron via weak processes 
\begin{eqnarray}
(\Omega \Omega )_{0^{+}\text{ }} &\rightarrow &(\Omega \Xi ^0)+\pi ^{-},
\label{w3} \\
(\Omega \Omega )_{0^{+}\text{ }} &\rightarrow &(\Omega \Xi ^{-})+\pi ^0,
\label{w4}
\end{eqnarray}
provided 
\begin{eqnarray}
E(2,6,-2) &>&E(2,5,-1)+m_{\pi ^{-}},  \label{w3e} \\
E(2,6,-2) &>&E(2,5,-2)+m_{\pi ^0},  \label{w4e}
\end{eqnarray}
respectively. On the other hand, the three-body weak decay channels for $%
(\Omega \Omega )_{0^{+}\text{ }}$ particle read 
\begin{eqnarray}
(\Omega \Omega )_{0^{+}\text{ }} &\rightarrow &\Omega +\Lambda +K^{-},
\label{b1} \\
(\Omega \Omega )_{0^{+}\text{ }} &\rightarrow &\Omega +\Xi ^0+\pi ^{-},
\label{c1} \\
(\Omega \Omega )_{0^{+}\text{ }} &\rightarrow &\Omega +\Xi ^{-}+\pi ^0.
\label{d1}
\end{eqnarray}
These channels can open provided the energy $E(2,|S|,Z)$ is large than the
sum of masses of the corresponding three particles.

We choose the following area 
\begin{eqnarray}
N_s-N_{\overline{s}} &\geq &0,  \label{25add} \\
Z &\geq &-A,  \label{26add} \\
|S|+Z &\leq &2A.  \label{27add}
\end{eqnarray}
to investigate the stability of dibaryon systems. By using the definition of
Schaffner-Bielich et. al \cite{8}, a dibaryon is called unstable if the
inequality (\ref{29}) is satisfied because in this case the strong decay
occurs. A dibaryon is called metastable if the inequality (\ref{29}) is not
satisfied but at least one of the weak decay inequalities is satisfied, in
this case the weak decay occurs. Other cases in which all inequalities be
broken are called absolutely stable.

Noting that the free energy $F$ reduces to the energy $E$ at zero
temperature, we can study the stability of dibaryon system by using the
formulas of Sec. II for QMDTD model. Our result is shown in Fig. 1 where
open circles stand for the unstable dibaryons and filled circles for the
metastable dibaryons. We find no dibaryon can be absolutely stable at zero
temperature. But three dibaryons, $(\Omega \Omega )_{0^{+}\text{ }},$ $%
(\Omega \Xi ^{-})$ and $(\Omega \Xi ^0)$ corresponding to $Q(2,6,-2),$ $%
Q(2,5,-2)$ and $Q(2,5,-1)$, respectively, are metastable. For $(\Omega
\Omega )_{0^{+}\text{ }}$ particle, its strong decay channel and a weak
channel (\ref{b1}) are forbidden, and the other weak channels, (\ref{a1})-(%
\ref{w3}), (\ref{c1}) and (\ref{d1}), are permitted. The above conclusions
are the same as that of Ref. \cite{z1,z3,z2}, though results of \cite
{z1,z3,z2} are established on the chiral SU(3) quark model and ours on the
QMDTD model.

\subsection{Binding energy and stable radius}

Secondly, we study the binding energy and the stable radius for above
metastable dibaryons. The binding energy of $(\Omega \Omega )_{0^{+}\text{ }}
$ is 
\begin{equation}
B(\Omega \Omega )_{0^{+}\text{ }}=2\times m_\Omega -E(\Omega \Omega )_{0^{+}%
\text{ }}.  \label{Binding energy}
\end{equation}
The stable radius $R$ of $(\Omega \Omega )_{0^{+}\text{ }}$ is determined by
the condition 
\begin{equation}
\frac{\delta E(\Omega \Omega )_{0^{+}\text{ }}}{\delta R}=0  \label{radius}
\end{equation}
at zero temperature. Obviously, the binding energy $E(\Omega \Omega )_{0^{+}%
\text{ }}$ depend on the values of the two parameters $B_0$ and $m_{s0}$.
The permitted ranges of $B_0$ and $m_{s0}$ are called ''stability window''
of the strange quark matter. We have calculated this stability window in
Ref. \cite{9} for QMDTD\ model. The binding energy and stable radius of $%
(\Omega \Omega )_{0^{+}\text{ }}$ state for fixed $m_{s0}=150$ $\mbox{MeV}$
and different $B_{0\text{ }}$are shown in Table 1: 
\begin{equation}
\begin{tabular}{||l|l|l|l||}
\hline\hline
$B_{0\text{ }}\mbox{MeV}$ $\mbox{fm}^{-3}$ & $m_{s0}$ $\mbox{MeV}$ & $%
B(\Omega \Omega )_{0^{+}\text{ }}$ $\mbox{MeV}$ & $R$ $\mbox{fm}$ \\ \hline
$170$ & $150$ & $162.7$ & $1.13$ \\ \hline
$175$ & $150$ & $143.7$ & $1.12$ \\ \hline
$180$ & $150$ & $125.2$ & $1.11$ \\ \hline
$182.5$ & $150$ & $116.1$ & $1.11$ \\ \hline
$190$ & $150$ & $89.2$ & $1.10$ \\ \hline\hline
\end{tabular}
.  \eqnum{Table 1}
\end{equation}
We find that the value of binding energy of $(\Omega \Omega )_{0^{+}\text{ }}
$ calculated by our model is in good agreement with that by chiral SU(3)
quark model. Hereafter we choose parameters $B_{0\text{ }}=182.5$ $\mbox{MeV}
$ $\mbox{fm}^{-3}$ and $m_{s0}=150$ $\mbox{MeV}$, because the binding energy 
$116.1$ $\mbox{MeV}$ is just equals to that predicted in Ref. \cite{z1}.
With these parameters, we draw the energy of $(\Omega \Omega )_{0^{+}\text{ }%
}$ state as a function of radius in Fig. 2. It is found the stable radius $%
R=1.11$ $\mbox{fm}$ is a little bigger than $0.84$ $\mbox{fm}$ given by
chiral SU(3) quark model.

To compare our result with chiral SU(3) quark model furthermore, we also
calculate the binding energy of $(\Omega \Xi ^{-})$ dibaryon 
\begin{equation}
B(\Omega \Xi ^{-})=m_\Omega +m_{\Xi ^{-}}-E(\Omega \Xi ^{-}).
\label{omiga-ksai BE}
\end{equation}
and its stable radius. The results are shown in Table 2: 
\begin{equation}
\begin{tabular}{||l|l|l|l||}
\hline\hline
$B_{0\text{ }}\mbox{MeV}$ $\mbox{fm}^{-3}$ & $m_{s0}$ $\mbox{MeV}$ & $%
B(\Omega \Xi ^{-})$ $\mbox{MeV}$ & $R$ $\mbox{fm}$ \\ \hline
$170$ & $150$ & $114.0$ & $1.13$ \\ \hline
$175$ & $150$ & $96.5$ & $1.12$ \\ \hline
$180$ & $150$ & $79.4$ & $1.12$ \\ \hline
$182.5$ & $150$ & $71.0$ & $1.11$ \\ \hline
$190$ & $150$ & $46.2$ & $1.10$ \\ \hline\hline
\end{tabular}
.  \eqnum{Table 2}
\end{equation}
We find that the values of binding energies of the $(\Omega \Xi ^{-})$ is
also consistent with the chiral SU(3) quark model \cite{z3}. The difference
between $(\Omega \Omega )_{0^{+}\text{ }}$ and $(\Omega \Xi ^{-})$ is about
several tens $\mbox{MeV}$. In particular, we hope to emphasize that the
QMDTD\ model gives the same conclusion as that of chiral SU(3) quark model:
the dibaryon $(\Omega \Omega )_{0^{+}\text{ }}$ is a deeply bound state, and
its binding energy is maximum in dibaryon systems.

\subsection{Mean lifetime}

The key for the detectability of the dibaryon is to study its mean life
time. Instead of the calculation of two-body decay Feynman diagram \cite{z1}%
, we employ a decay formula which is first given by Chin and Kerman \cite
{qing} to estimate the mean life time of $(\Omega \Omega )_{0^{+}\text{ }}$
and $(\Omega \Xi ^{-})$ via weak decay. The decay formula is 
\begin{equation}
1/\tau =[G^2\mu _s^5/192\pi ^3]\sin ^2\theta _cF(z),  \label{qing}
\end{equation}
with 
\begin{eqnarray}
F(z) &=&1-8z+8z^3-z^4-12z^2\ln z,  \label{fz} \\
z &=&\mu _u^2/\mu _s^2,  \label{z}
\end{eqnarray}
where the Cabibbo angle is given by $\sin \theta _c\simeq 0.22$. With Eqs. (%
\ref{qing})-(\ref{z}), we can roughly estimate order-of-magnitude of the
dibaryon's mean life time. The coupling constant $G$ is determined as
follows: using Eq. (\ref{qing}) to calculate the decay mean lifetime of $%
\Omega $ for following channels: 
\begin{eqnarray}
\Omega  &\rightarrow &\Lambda +K^{-},  \label{o1} \\
\Omega  &\rightarrow &\Xi ^0+\pi ^{-},  \label{o2} \\
\Omega  &\rightarrow &\Xi ^0+\pi ^0  \label{o3}
\end{eqnarray}
and fit the value with the $\Omega $ mean lifetime $8.21\times 10^{-11}$ $%
\mbox{s}$ to determine coupling constant $G$. Then use this coupling
constant $G$ to calculate the lifetime of the $(\Omega \Omega )_{0^{+}\text{ 
}}$ and $(\Omega \Xi ^{-})$. With the parameters $B_{0\text{ }}=182.5$ $%
\mbox{MeV}$ $\mbox{fm}^{-3}$, $m_{s0}=150$ $\mbox{MeV}$, we obtain the mean
lifetime is $4.7\times 10^{-11}$ $\mbox{s}$ for $(\Omega \Omega )_{0^{+}%
\text{ }}$ and $6.2\times 10^{-11}$ $\mbox{s}$ for $(\Omega \Xi ^{-})$,
respectively.

\section{Result at finite temperature}

Now we turn to discuss the thermodynamic properties, especially, the
stability of dibaryons at finite temperature. Since according to the chiral $%
SU(3)$ quark model and QMDTD model, the best stable candidate of dibaryons
is $(\Omega \Omega )_{0^{+}\text{ }}$ at zero temperature, we focus our
attention only on $(\Omega \Omega )_{0^{+}\text{ }}$in this section. Instead
of the energy at zero temperature, we calculate the free energy density
first. Given the strangeness number, baryon number and electric charge, the
free energy and stable radius for dibaryons and baryons can be obtained
self-consistently by using the formulas in Sec. II.

As was pointed out by \cite{z1,z3,z2} and sec. III, the strong decay channel 
\begin{equation}
(\Omega \Omega )_{0^{+}\text{ }}\rightarrow \Omega +\Omega ,
\label{twoomigas}
\end{equation}
is forbidden because the binding energy of $(\Omega \Omega )_{0^{+}\text{ }} 
$is large. But at high temperature, a metastable dibaryon can absorb enough
energy from the hot environment to overcome its binding energy and open the
strong decay channel. To study the possibility of strong decay, we treat the 
$(\Omega \Omega )_{0^{+}\text{ }}$and $\Omega $ as a six quarks cluster with 
$Q(2,6,-2)$ and a three quarks cluster with $Q(1,3,-1)$ respectively, and
calculate their free energy density. Define 
\begin{equation}
\Delta f=f(\Omega \Omega )_{0^{+}\text{ }}-2f(\Omega )  \label{deltaf}
\end{equation}
as the difference between the free energy density of $(\Omega \Omega )_{0^{+}%
\text{ }}$and two separate $\Omega $ system, the curve of $\Delta f$ vs.
temperature $T$ is shown in Fig. 3. We find $\Delta f$ increases with
temperature. It becomes zero when temperature arrives at $T_0=129.3$ $%
\mbox{MeV}$. When $T>T_0$, $\Delta f>0$, the strong decay channel opens.
This result is reasonable if we notice that the temperature $T_0$ is a
little bigger than the binding energy of $(\Omega \Omega )_{0^{+}\text{ }}$, 
$116.1$ $\mbox{MeV}$.

Since the strong decay of $(\Omega \Omega )_{0^{+}\text{ }}$will open and $%
(\Omega \Omega )_{0^{+}\text{ }}$becomes unstable when $T>T_0$, the
favourable detectability of $(\Omega \Omega )_{0^{+}\text{ }}$must be taken
in the regions $0\leq T\leq 129.3$ $\mbox{MeV}$. This temperature regions
seem wide enough for detectability in RHIC experiments. Of course, in order
to help the design of a detectable experiment, it is necessary to get more
information about $(\Omega \Omega )_{0^{+}\text{ }}$in this metastable
regions. For this purpose, we study the temperature dependence of the radius
and the mean lifetime of $(\Omega \Omega )_{0^{+}\text{ }}$in $0\leq T\leq
T_0$ regions.

The curves of stable radius $R$ vs. $T$ for $(\Omega \Omega )_{0^{+}\text{ }%
} $is shown in Fig. 4. We see that the radius increases from $1.11$ $%
\mbox{fm}$ to $1.89$ $\mbox{fm}$ when temperature increases from $0$ to $%
129.3$ $\mbox{MeV}$. the volume of the bag expands when temperature
increases.

Finally, we still use Eqs. (\ref{qing})-(\ref{z}) to study the temperature
effect on the mean lifetime. Instead of Fermi energies of quarks at zero
temperature in these equations, we use the temperature dependent chemical
potentials $\mu _u(T)$ and $\mu _s(T)$ to estimate the order of magnitude
for the mean lifetime of $(\Omega \Omega )_{0^{+}\text{ }}$. We find that
the order of magnitude is not less than $10^{-11}$ $\mbox{s}$ in the whole
metastable temperature regions.

\section{ Summary}

In summary, we have studied thermodynamic properties of dibaryons by using
QMDTD model. At zero temperature, we find that only three dibaryons, namely, 
$(\Omega \Omega )_{0^{+}\text{ }}$, $(\Omega \Xi ^{-})$ and $(\Omega \Xi ^0)$%
, are metastable. At least one weak decay channel opens for these states and
all the strong decay channels close. We calculate the binding energy, stable
radius as well as the mean lifetime of $(\Omega \Omega )_{0^{+}\text{ }}$and 
$(\Omega \Xi ^{-})$, it is found that our results are in good agreement with
those given by chiral $SU(3)$ quark model. Using two essentially different
models and get the same conclusion, this result strongly compels us to
support $(\Omega \Omega )_{0^{+}\text{ }}$is the most favorable candidate of
dibaryon for detectability, because its binding energy is maximum.

Extending our investigation to finite temperature, we find the strong decay
channel will open and $(\Omega \Omega )_{0^{+}\text{ }}$will become unstable
when temperature arrives at $T_0=129.3$ $\mbox{MeV}$. We also calculate the
radius and the lifetime of $(\Omega \Omega )_{0^{+}\text{ }}$at finite
temperature.

\section{Acknowledgment}

We thank professors Y. W. Yu, Z. Y. Zhang, T. H. Ho, C. R. Ching and P. N.
Shen for helpful discussions, we also thank professor F. Wang for sending us
his reprints. This work was supported in part by the NNSF of China under
contract Nos. 10047005, 19947001 and 10235030.

\section{Figure Captions}

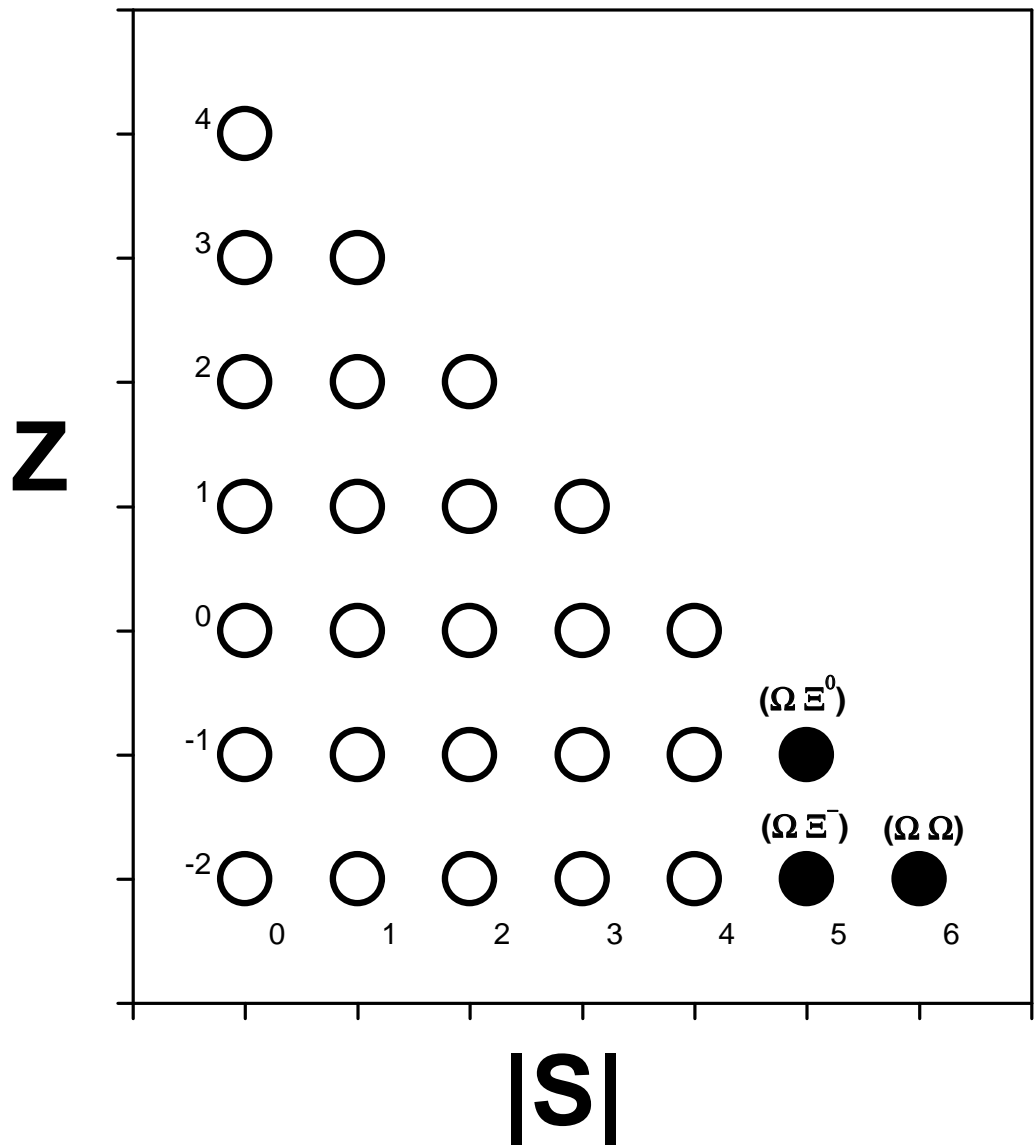
Figure 1. The electric charge $Z$ as a function of the strangeness number $%
|S|$ for unstable strangelets (open circles), metastable strangelets (filled
circles) with baryon number $A=2$ at zero temperature.

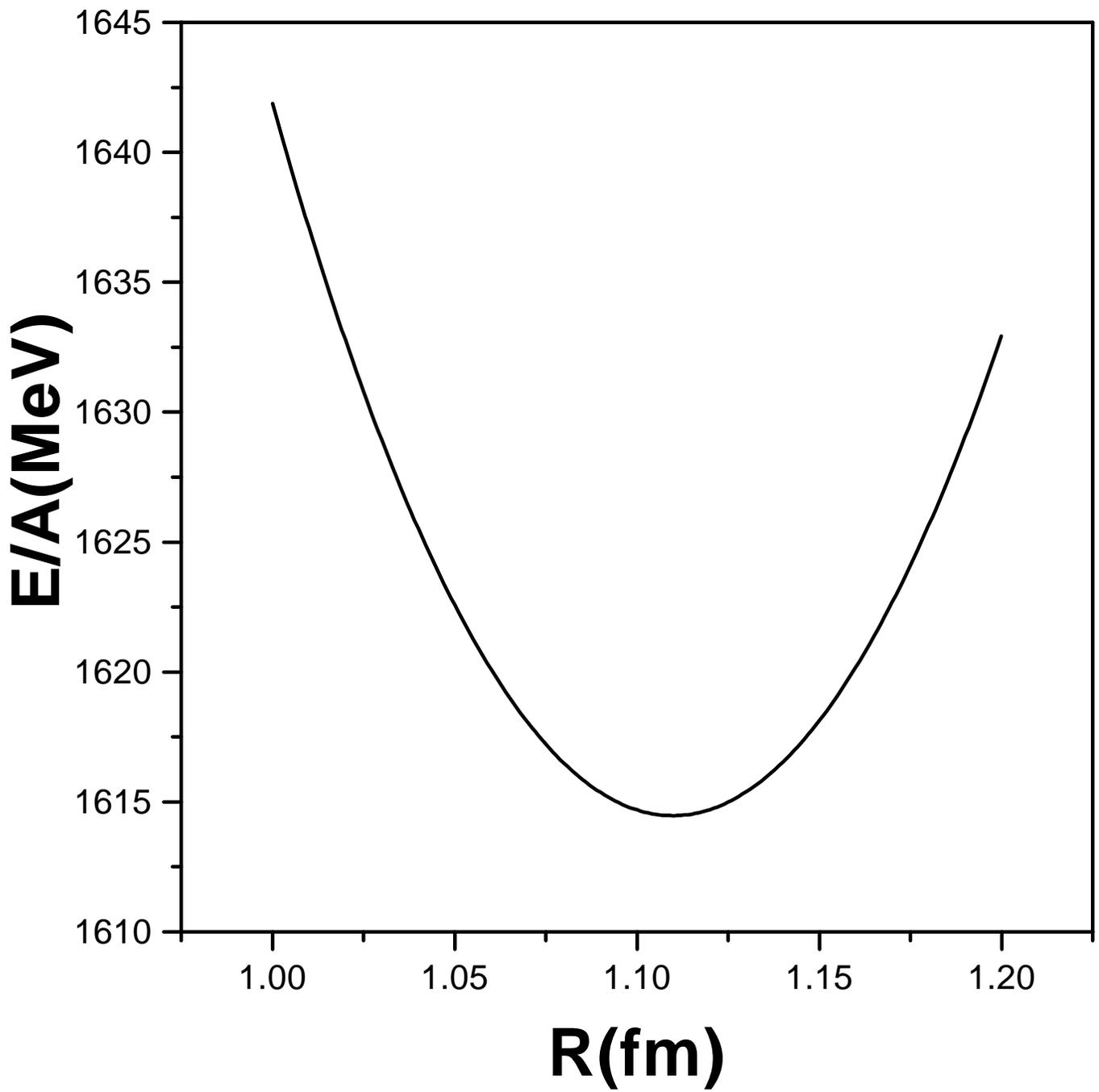
Figure 2. The energy per baryon number $E/A$ as a function of the radius $R$
for $(\Omega \Omega )_{0^{+}\text{ }}$particle.

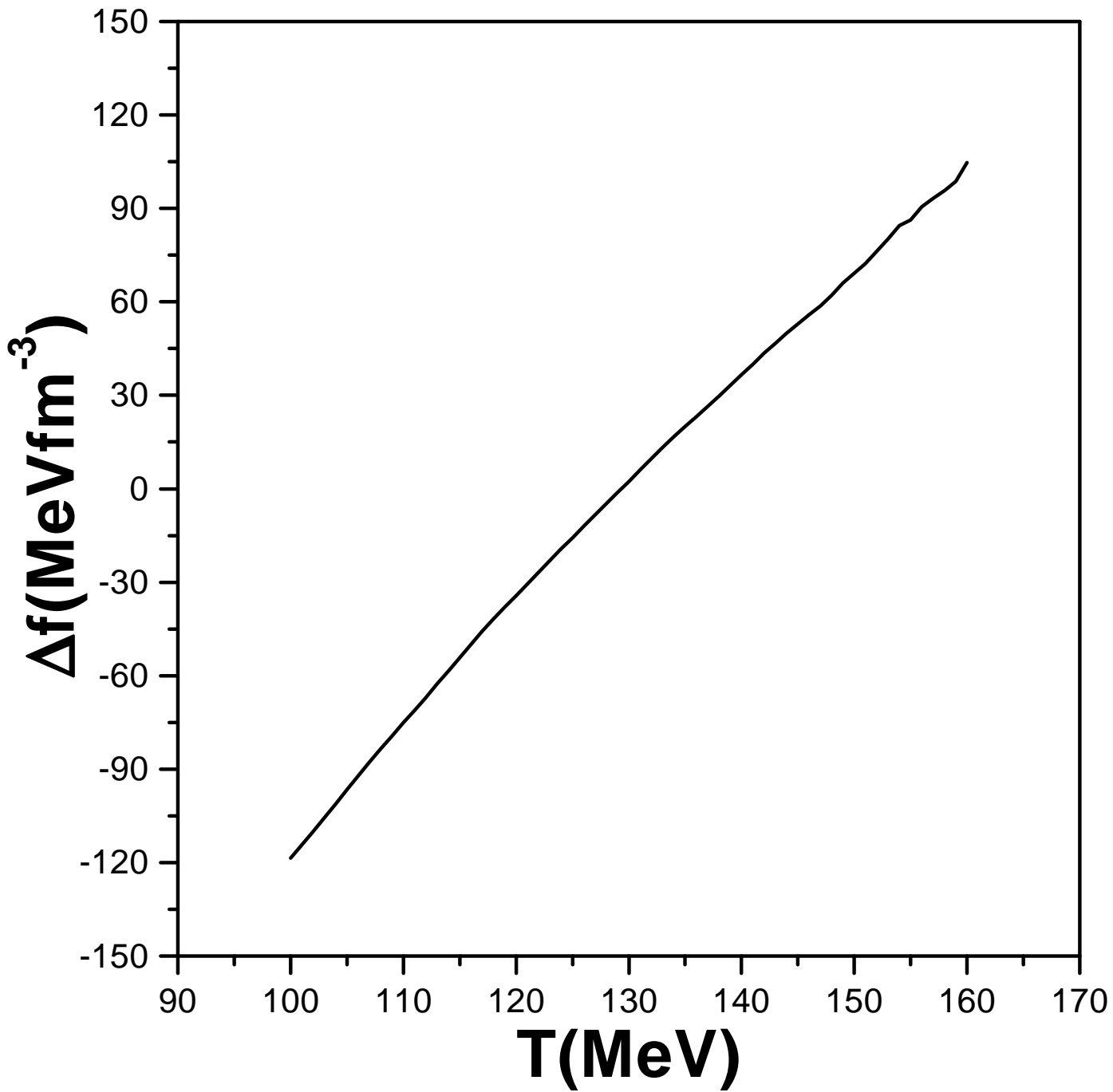
Figure 3. The difference of free energy density between $(\Omega \Omega
)_{0^{+}\text{ }}$ particle and two free $\Omega $ particles $\Delta f$ as a
function of temperature $T$.

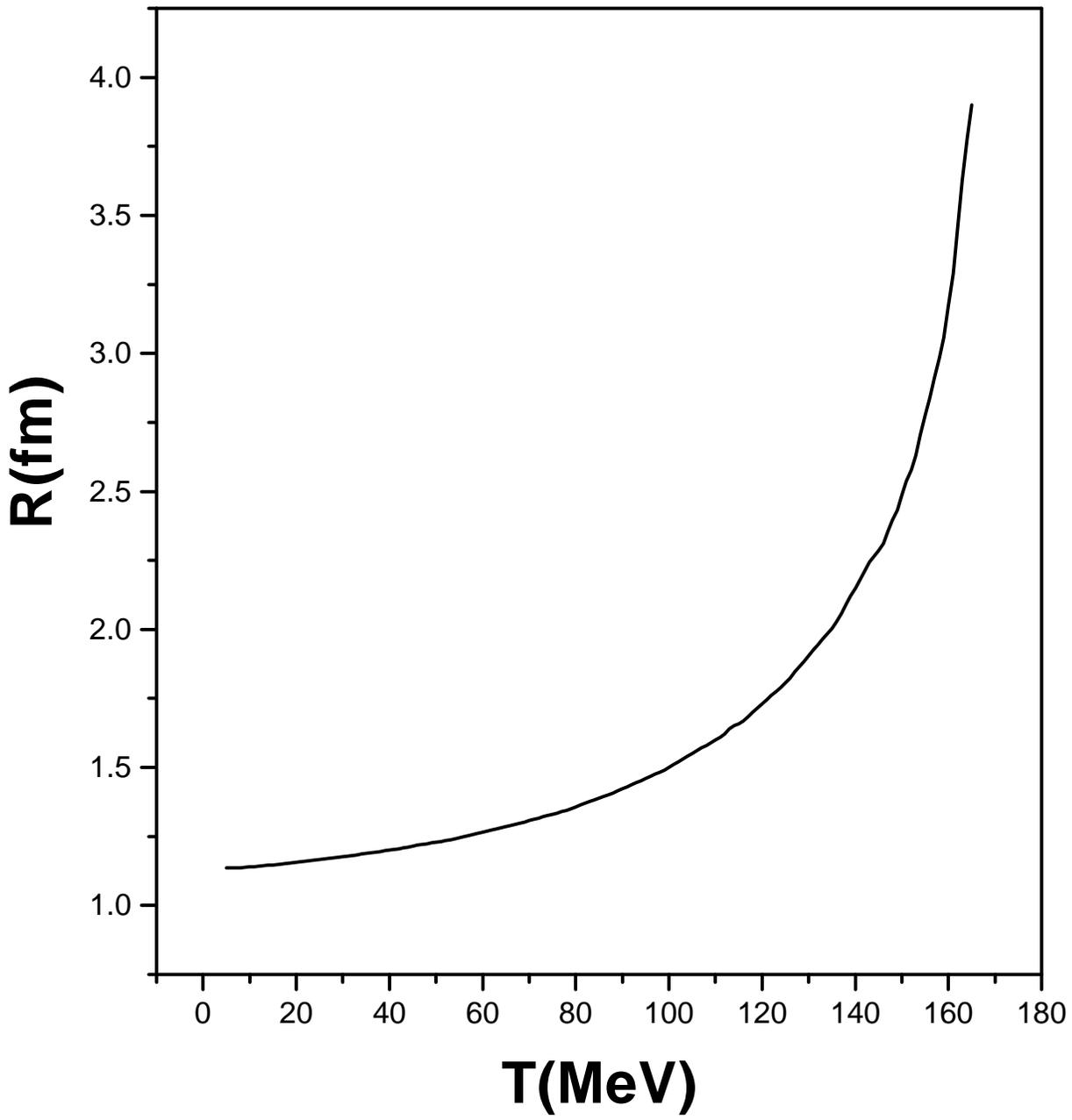
Figure 4. The stable radius $R$ as a function of the temperature $T$ for $%
(\Omega \Omega )_{0^{+}\text{ }}$particle.

\section{Table Captions}

Table 1. The binding energy and stable radius of $(\Omega \Omega )_{0^{+}%
\text{ }}$particle.

Table 2. The binding energy and stable radius of $(\Omega \Xi ^{-})$
particle.

\end{document}